\documentclass[a4paper]{jpconf}
\usepackage{graphicx}
\begin{document}
\title{The CONNIE experiment}

\author{
A.~Aguilar-Arevalo$^1$,
X.~Bertou$^2$,
C.~Bonifazi$^3$,
M.~Butner$^4$, 
G.~Cancelo$^4$, 
A.~Castaneda~Vazquez$^1$, 
B.~Cervantes~Vergara$^1$, 
C.R.~Chavez$^5$, 
H.~Da~Motta$^6$, 
J.C.~D'Olivo$^1$, 
J.~Dos Anjos$^6$, 
J.~Estrada$^4$, 
G.~Fernandez Moroni$^{7,8}$, 
R.~Ford$^4$, 
A.~Foguel$^{3,6}$, 
K.P.~Hernandez Torres$^1$, 
F.~Izraelevitch$^4$, 
A.~Kavner$^9$,
B.~Kilminster$^{10}$, 
K.~Kuk$^4$, 
H.P.~Lima Jr.$^6$, 
M.~Makler$^6$, 
J.~Molina$^5$, 
G.~Moreno-Granados$^1$, 
J.M.~Moro$^{11}$, 
E.E.~Paolini$^{7,12}$, 
M.~Sofo Haro$^2$, 
J.~Tiffenberg$^4$, 
F.~Trillaud$^1$, 
and S.~Wagner$^{6,13}$
}

\address{
$^1$Universidad Nacional Aut\'onoma de M\'exico, Ciudad de M\'exico, M\'exico \\
$^2$Centro At\'omico Bariloche - Instituto Balseiro, CNEA/CONICET, Argentina \\
$^3$Universidade Federal do Rio de Janeiro, Instituto de F\'isica, Rio de Janeiro, Brazil \\
$^4$Fermi National Accelerator Laboratory, Batavia, IL, U.S.A. \\
$^5$Facultad de Ingenier\'ia - Universidad Nacional de Asunci\'on, Paraguay \\
$^6$Centro Brasileiro de Pesquisas Fisicas, Rio de Janeiro, Brazil \\
$^7$Departamento de Ingeniería El\'ectrica y de Computadores, Universidad Nacional del Sur, Bah\'ia Blanca, Argentina \\
$^8$Instituto de Investigaciones en Ingenier\'ia El\'ectrica ``Alfredo Desages'', CONICET - Universidad Nacional del Sur, Bah\'ia Blanca, Argentina \\
$^9$University of Michigan, Ann Arbor, MI, U.S.A. \\
$^{10}$Universit\"at Z\"urich Physik Institut, Zurich, Switzerland \\
$^{11}$Depto. de Ingenier\'ia, Universidad Nacional del Sur, Bah\'ia Blanca, Argentina \\
$^{12}$Comisi\'on de Investigaciones Cient\'ificas Provincia Buenos Aires, La Plata, Argentina \\
$^{13}$Pontificia Universidade Cat\'olica, Rio de Janeiro, Brazil
}

\ead{alexis@nucleares.unam.mx}

\begin{abstract}
The CONNIE experiment uses fully depleted, high resistivity CCDs as particle detectors in an attempt to measure for the first time the Coherent Neutrino-Nucleus Elastic Scattering of antineutrinos from a nuclear reactor with silicon nuclei.
%The low energy threshold (of order ~10 eV), and the relatively large mass of the CONNIE CCDs make them an ideal detector technology to observe this process, which is predicted by the Standard Model. In 2014 a prototype detector was installed 30 m from the core of one of the reactors at the ANGRA nuclear power plant in Brazil, where the characterization of the detectors response to the backgrounds is being studied. 
This talk, given at the XV Mexican Workshop on Particles and Fields (MWPF), discussed the potential of CONNIE to perform this measurement, the installation progress at the Angra dos Reis nuclear power  plant, as well as the plans for future upgrades.

\end{abstract}

\vspace{-0.3cm}

\section{Introduction}

The Coherent Elastic Neutrino-Nucleus Scattering (CE$\nu$NS), is a Standard Model (SM) process where a neutrino, or antineutrino, interacts with a nucleus as a whole entity \cite{freedman:1974}. It arises from the coherent enhancement of the interaction cross-section with the constituent nucleons, when the 4-momentum transfer is small compared to the reciprocal of the nuclear size: $|q^2|<1/R^2$; in the laboratory frame, this corresponds roughly to incident neutrino energies $E_\nu<50$~MeV.
Its differential cross-section, to lowest order in $T/E_{\nu}$, is \cite{barranco:2005}

\begin{equation}
\label{xsec}
\frac{d\sigma}{dT}(E_{\nu},T) = \frac{G_F^2}{8\pi} \; [Z(4\sin^2\theta_W-1)+N]^2 \;
M \left(2 - \frac{MT}{E_{\nu}^2} \right) |f(q^2)|^2 \;,
\end{equation}

\noindent
where $T$ is the nuclear recoil energy, $G_F$ is the Fermi constant, $M$ is the mass of the nucleus,  $Z$, and $N$ are the number of protons and neutrons in the nucleus respectively, $\theta_W$ is the weak mixing angle, and $f(q^2)$ is the nuclear form factor, which is a function only of the 4-momentum transfer $q^2= 2E_\nu TM/(T-E_\nu)$. 
Integration of Eq.~(\ref{xsec}) over $T$ with the approximations $f(q^2)\approx 1$ (valid for $E_{\nu}<20~{\rm MeV}$), and $\sin^2\theta_W\approx1/4$,  yields a total cross section as a function of neutrino energy: 
$\sigma_T(E_{\nu})\approx 4.22\times10^{-45}N^2(E_{\nu}/1~{\rm MeV})^2\:{\rm cm}^2$.

This process has never been observed experimentaly, primarily because, in addition to its small cross-section (${\cal O}\sim10^{-41}~{\rm cm}^2$), the nuclear recoil energies involved are very small ($<\sim$15 keV), and escape the capabilities of most sufficiently massive detectors. 
Many detector technologies used in direct WIMP dark matter (DM) searches, combining low thresholds with large target masses, have been considered for the detection of CE$\nu$NS. Such DM experiments are expected to eventually reach the sensitivity to the ``neutrino floor'': an irreducible background from the coherent scattering of solar, atmospheric and difuse supernova neutrinos \cite{cushman:2013}.

Recent developments in the technology of Charge Coupled Devices (CCD) have made it possible to build and operate detectors with very low thresholds (of order $\sim$10 eV) and relatively large sensitive mass ($\sim$5~g per CCD); these have been successfully used in the DAMIC experiment \cite{damic} at SNOLAB to perform direct DM searches in the low mass region $<10$~GeV$/c^2$, demostrating the unique capabilities of CCDs for the detection of low-energy recoils.

The Coherent Neutrino Nucleus Interaction Experiment (CONNIE) is a detector of solid-state technology installed at a nuclear reactor with an array of CCDs at its core, designed to observe the CE$\nu$NS reaction for the first time. In 2014 a prototype detector was installed at one of the reactors at the Angra dos Reis nuclear power plant in Brazil, where the characterization of the detectors response to the backgrounds is being studied.

\section{The CONNIE CCDs}

\begin{figure}[t]
\centering
\scalebox{0.4}{\includegraphics{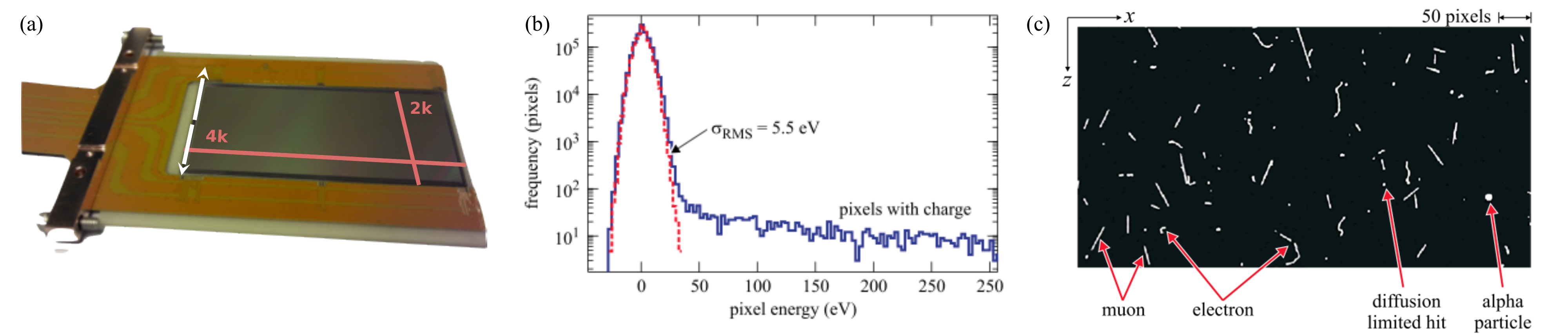}}
\caption{\small (a) A packaged 8~Mpix CONNIE CCD. (b) Single pixel value distribution at -140$^\circ$C with and without a $^{60}$Co gamma source. (c) Sea level exposure showing various particle tracks.}
\label{fig-ccd}
\end{figure}

The CCDs are developed by the LBNL Microsystems Laboratory. They are fabricated on high-resistivity (10-20~k$\Omega\;$cm) silicon, which allows their operation at full depletion with moderate substrate voltages. The sensitive area is divided into several million square (15~$\mu$m$\times$15~$\mu$m) pixels, reaching a few tens of square centimeters, and thicknesses up to $\sim650~\mu$m, providing detector masses of a few grams per CCD. They exhibit very low charge transfer inefficiencies and dark current when operated at low temperatures ($\sim140$~K). The energy response to electronic and nuclear recoils has been extensively studied \cite{elec-rec-calib-1,elec-rec-calib-2,nuc-rec-calib-1,nuc-rec-calib-2}. The CCDs are packaged in supporting structures and bonded to flex readout cables, as shown in Figure~\ref{fig-ccd}~(a), at the Fermilab Silicon Detector Facility.
An optimal readout noise of $\sigma_{\rm RMS}=1.5{\rm e}^-$ (equivalent to $\sim$5.8~eV) on the pixel values can be achieved by appropriate choice of the output amplifier integration time window. Figure~\ref{fig-ccd}~(b) shows the distribution of pixel values in a typical exposure with a $^{60}$Co source.
A typical image from a CCD exposure at sea level in which different particle tracks are visible is shown in Figure~\ref{fig-ccd}~(c).
%short curved tracks are typical of energetic electrons from electromagnetic interactions; straight tracks are produced by cosmogenic muons; large circular structures are produced by alpha particles. 
Events depositing energy within the volume of a single pixel produce point-like events where ionized charge diffuses over one or a few neighboring pixels as it is drifted towards the CCD gate electrodes. These ``diffusion limited hits'' are expected from nuclear recoils such as those produced by the coherent neutrino-nucleus scattering.

\section{Prototype installation at Angra-II}

The prototype CCD array in the initial installation (2014 to mid 2016) has four 4k$\times$2k pixel (8~Mpix) CCDs, each with a thickness of 250~$\mu$m, an area of 6~cm$\times$3~cm, and a mass of $\sim 1$~g. These are "engineering grade" sensors having several defects and significant contamination from U and Th in the AlN substrate used in the package design; they were deployed with the purpose of testing the system operation at the reactor site, and have preliminary estimates of the environmental background. Only two of the sensors were of sufficient quality for the background analysis.

\begin{figure}[t]
\centering
\scalebox{0.49}{\includegraphics{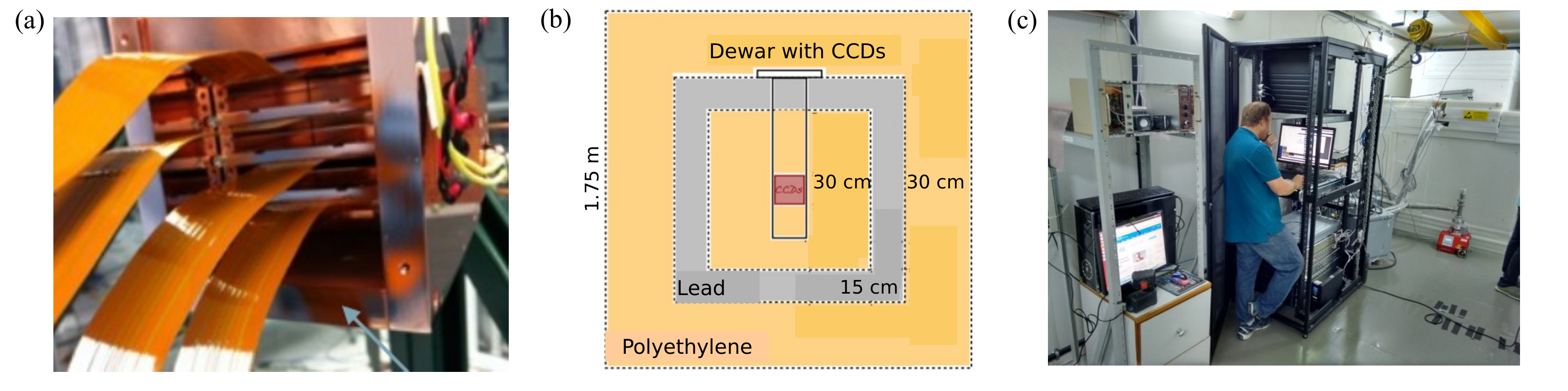}}
\caption{\small (a) CCD copper box. (b) Full shield concept. (c) Full installation inside container.}
\label{fig-det}
\end{figure}

The CCDs are stacked in a high purity copper box, see Figure~\ref{fig-det}~(a) kept at a temperature of $-140^\circ$C inside a vacuum vessel ($10^{-7}$~torr). A 15~cm-long lead cylinder inside the vessel shields the CCD array from gammas coming from radioactive contaminants in the readout electronics.
A radiation shield was installed in two phases: Phase-I consisting of an inner 30~cm layer of polyethylene followed by 5~cm of lead, and Phase-II consisting of an additional 10~cm of lead (totaling 15~cm of Pb) and an outer 30~cm layer of polyethylene. The concept of the full shield is shown in Figure~\ref{fig-det}~(b). A picture of the inside of the container with the final installation of the detector, electronics, and support systems is shown in Figure~\ref{fig-det}~(c).

The detector is located at a distance of 30~m, see Figure~\ref{fig-reactor}~(a) and (b), from the core of the Angra-II reactor of the Almirante Alvaro Alberto nuclear power plant in Angra dos Reis, near Rio de Janeiro, Brasil. It is housed in a standard shipping container outside the main reactor building. The Angra-II reactor is a 3.95~GW$_{\rm th}$ Presurized Water Reactor, and generates $\sim 8.7\times10^{20}$~$\bar\nu_e$s$^{-1}$. At the detector location this corresponds to $\sim 7.8\times10^{20}\;\bar\nu_e$cm$^{-2}$s$^{-1}$, coming mainly from $\beta$ decay chains of the products of the fragmentation of fissile isotopes $^{235}$U, $^{238}$U, $^{239}$Pu, and $^{241}$Pu, and n-capture processes in the reactor core (avg. fission rate of $\sim1.2\times10^{20}$fis/s). The antineutrino flux expected from the reactor is shown in Figure~\ref{fig-reactor}~(c).

\section{Expected event rates and forecast}

\begin{figure}[t]
\centering
\scalebox{0.41}{\includegraphics{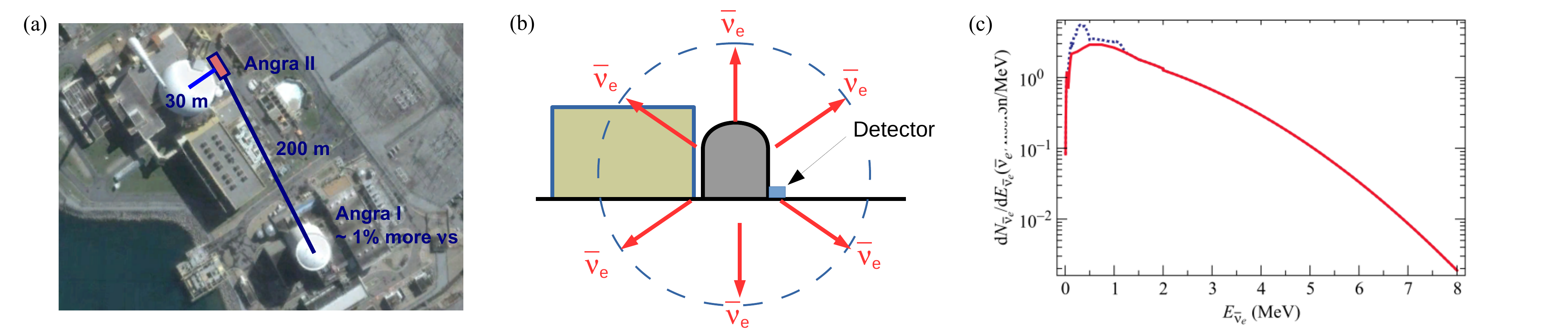}}
\caption{\small (a) Detector location in reactor complex. (b)  Detector scketch. (c) Neutrino flux at detector.}
\label{fig-reactor}
\end{figure}

\begin{figure}[t]
\centering
\scalebox{0.6}{\includegraphics{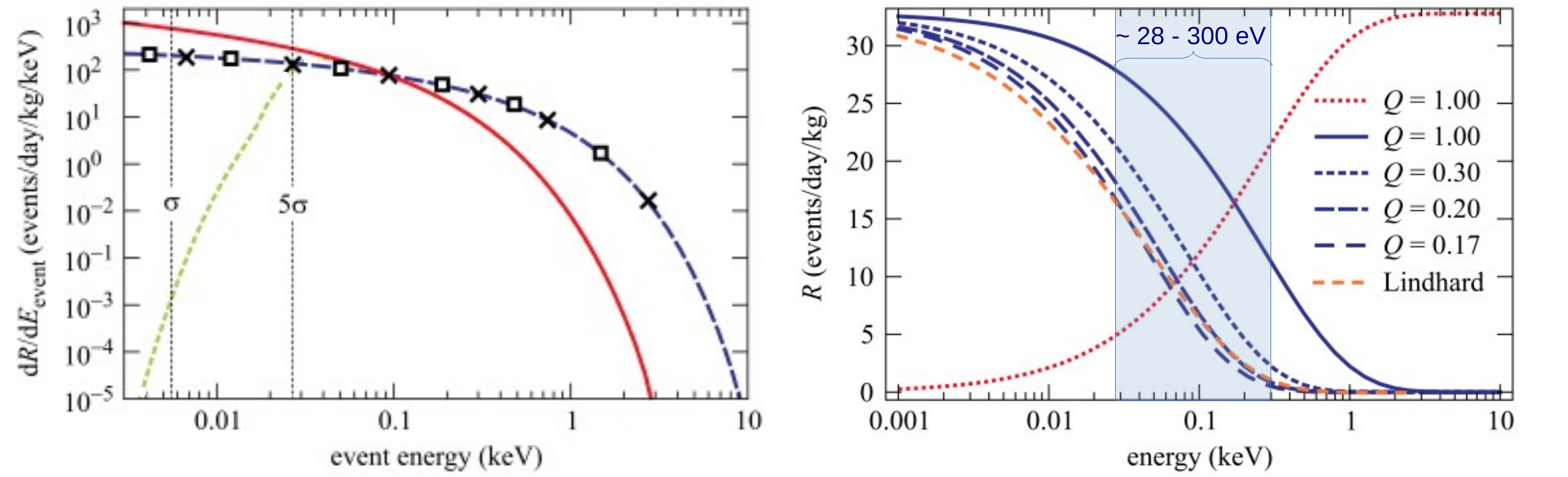}}
\caption{\small 
Left: The expected recoil energy spectrum (blue dashed line); the visible energy spectrum assuming a Lindhard $Q$ factor (red solid line); the visible energy spectrum with Lindhard $Q$ factor and the effect of the detector efficiency (green dashed line).
Right: Total event rate as a function of the energy threshold for various quenching factors. The shaded area shows the expected signal region for a threshold of $5\sigma_{RMS}\simeq 28$~eV, and realistic quenching factors.
Adapted from \cite{guillermo:2015}.
}
\label{fig-rates}
\end{figure}

Convolution of the antineutrino energy spectrum in Figure~\ref{fig-reactor}~(c), with the differential cross section in Eq.~(\ref{xsec}) yields the nuclear recoil energy spectrum. This is converted into the measured spectrum by means of the quenching factor $Q(T)$, relating the ``measured'' energy, $T_{m}$, to the recoil energy, $T$, via $T_{m}=T\;Q(T)$. Event rates are then calculated as a function of the detection threshold, $T_{\rm thr}$, by integration of the measured energy spectrum weighted by the detection efficiency.
%\begin{equation}
%R(T_{\rm thr}) = \int_{T_{\rm thr}}^{\infty}\; dT_{v}\;\frac{dR}{dT_{v}}(T_{v}) \times {\cal E}(T_{v})
%\end{equation}
%\noindent
%where ${\cal E}(T_{v})$ is the detection efficiency. For the estimates presented here the efficiency was determined from a simple threshold algorithm making use of the charge distribution pattern and an enhanced signal-to-noise ratio by configuring the CCD readout in $10\times10$ pixel binning 
%
Details of the event rate calculations and the efficiency estimate assumed here are given elsewhere \cite{guillermo:2015}.
The left-hand side plot in Figure~\ref{fig-rates} compares the expected recoil energy spectrum and the spectrum of measured energies assuming the Lindhard \cite{lindhard:1963} quenching factor with (green dashed line) and without (red dashed line) the effect of the detector efficiency. The right-hand side plot shows the total event rate as a function of the threshold, $T_{\rm thr}$, for different choices of $Q$ showing a weak dependence on the details of the quenching factor.
From these calculations a signal rate of 16.15~evt/kg/day is extracted, using the Lindhard quenching factor and the estimated efficiency. As shown in the right-hand side plot, the contribution to the total number of signal events above 300~eV is negligible, hence defining a signal region from $\sim28-300$~eV.
Earlier studies have shown that with passive shielding at shallow depths, a background rate at sea level of $\sim 600$~evt/day/keV/kg can be achieved  \cite{heusser:1995}; over the expected signal region this amounts to a background contribution of 163.2~evt/day.

\subsection{Forecast for a 52~g detector}

For the purpose of evaluating the potential of CONNIE, in what follows we assume a 52~g detector array (ten 8~Mpix CCDs with 650~$\mu$m and fiducial mass of 5.2~g each) running for $T$ days. As shown in \cite{guillermo:2015}, in such a detector we expect a signal rate of $\sim 0.84 T$ events, and a background rate of $\sim 8.5 T$~events. Adding a rate of false positive events of $\sim$3.8~evt/day due to readout noise values above the threshold gives a total of $\sim 11.68 T$ background events. The significance of a given signal can be calculated as a function of the running time by the expression 
$N_\sigma = 0.84\;T / \sqrt{11.68\;T} = 0.25 \sqrt{T}$. 
Table \ref{significances} shows the running time in days required to reach a given level of significance with this setup. A $\sim 3 \sigma$ significance can be achieved with 150 live days under the above assumptions.

\begin{table}[h]
\centering
\caption{\small Days to reach a given confidence level (C.L.).}
\begin{tabular*}{9cm}{@{\extracolsep{\fill}}l|ccccc} \hline
C.L. [\%] & 80 & 90 & 95 & 98 & 99  \\
T (days)  & 12 & 28 & 45 & 70 & 150 \\ \hline
\end{tabular*}
\label{significances}
\end{table}

\section{Timeline}

Plans to deploy a CCD-based detector making use of the available space already aproved for the Angra-Neutrinos project at the Angra-II reactor site began in 2013. 
The detectors were tested and packaged at Fermilab during 2014.
The fully assembled vacuum vessel (containing the CCDs), the cooling and vacuum systems, and the readout electronics, were all shipped from Fermilab to Rio de Janeiro on August 2014, arriving (by sea) on September. 
The installation of the system took place from October through November; producing the first images of events for background studies as early as December.
The full (lead+polyethylene) shield was assembled during July and August of 2015. 
From August through September, 2015 more than a month of reactor-on data were acquired with the full shield in place. 
From September through October, 2015, one full month of reactor-off data were collected during a scheduled refueling outage. 
These data sets constitute the first CONNIE engineering run, whose results have been published since the time of this conference \cite{1st:connie:paper:2016}.

\section{Summary and outlook}

The low noise and low threshold of CCDs make them an ideal detector technology for the measurement of low energy depositions such as those expected from the coherent scattering of neutrinos off nuclei. The CONNIE Collaboration has been operating a prototype CCD detector at the Angra-II reactor in Brazil since December of 2014, and has taken data with different shield configurations, and with the reactor on and off. Results have been published since the XV~MWPF \cite{1st:connie:paper:2016} demonstrating the feasibility of the operations and background stability. The current setup is not expected to distinguish the coherent neutrino scattering events, only to demonstrate operations at the reactor site and measure environmental backgrounds and stability. Plans for upgrade to a CONNIE100 are ongoing.

\subsection{Update since the XV~MWPF}

The collaboration plans to conduct an upgrade to a ∼100 g detector (CONNIE100) during 2016. The plan includes an upgrade of the packaging design that will reduce the intrinsic background of the detector ({\it e.g.} removal of the AlN substrate). It will incorporate a total of 18 engineering grade sensors of superior quality, with 675~$\mu$m thickness and 5.7~g of active mass per CCD. The upgraded detector should achieve sensitivity to the SM neutrino-nucleus coherent scattering as discussed in
\cite{guillermo:2015}.

\section*{Acknowledgements}
We thank {\it Central Nuclear  Almirante Alvaro Alberto Electronuclear} for access to the reactor and their technical support, in particular to Ilson Soares.
We thank the Silicon Detector Facility team at Fermilab for being the host lab for the assembly and testing of the detector components.
%
%This work is partially supported by the U.S. Department of Energy, Office of Science, Office of High Energy Physics. Fermi National Accelerator Laboratory is operated by Fermi Research Alliance, LLC under Contract No. De-AC02-07CH11359 with the United States Department of Energy.
%
We acknowledge support from Minist\'erio da Ciencia, Tecnologia e Inova\c{c}\~ao (MCTI) and the Brazilian funding agencies FAPERJ (grant E-26/110.145/2013), CNPQ, and FINEP;  M\'exico's CONACYT (grant No. 240666), and DGAPA-UNAM (PAPIIT grants IB100413 and IN112213); Argentina's Agencia Nacional de Promoci\'on Cient\'ifica y Tecnol\'ogica (PICT-2014-1225, PICT-2013-2128). We thank Ricardo Galv\~ao and Ronald Shellard for supporting the experiment.
%
%We thank Ricardo Galvo and Ronald Shellard for their support to the experiment.

\end{document}